\documentclass[12pt]{article}
 \usepackage[dvips]{graphicx}\usepackage{amsmath}
 \begin{document}

\title{The Programs of the Extended Relativity in C-spaces: Towards the
Physical
Foundations of String
Theory }
\author{C. Castro\thanks{Center for Theoretical Studies of Physical Systems,
Clark
Atlanta University, Atlanta GA. 30314;~~  Email:
castro@ctsps.cau.edu}}
\date{May 2002}
\maketitle
\begin{abstract}
An outline is presented  of the Extended Scale Relativity (ESR)
in C-spaces (Clifford manifolds), where the speed of light and
the minimum Planck scale are the two universal invariants. This
represents in a sense an extension of the theory developed by
L.Nottale long ago. It is shown how all the dimensions of a
C-space can be treated on equal footing by implementing the
holographic principle associated with a nested family of p-loops
of various dimensionalities. This is achieved by using poly-vector
valued coordinates in C-spaces that encode in one stroke points,
lines, areas, volumes,... In addition, we review the derivation of
the minimal-length string uncertainty relations; the logarithmic
corrections (valid in any dimension) to the black hole
area-entropy relation. We also show how the higher derivative
gravity with torsion and the recent results of kappa-deformed
Poincare theories of gravity follow naturally from the geometry of
C-spaces. In conclusion some comments are made on the cosmological
implications of this theory with respect to the cosmological
constant problem, the two modes of time, the expansion of the
universe, number four as the average dimension of our world and a
variable fine structure constant.
\end{abstract}

\section{Introduction}

Since the inception of string theory there have been an incessant
strive to find the underlying fundamental physical principle
behind string theory, not unlike the principle of equivalence and
general covariance in Einstein's general relativity. This
principle might well be  related to the existence of an invariant
minimal length scale (Planck scale) attainable in nature. In this
connection it should be said that a deeper understanding of
geometry and its relation to algebra has always turned out be very
useful for the advancement of physical theories. Without
analytical geometry Newton mechanics, and later special
relativity, could not have acquired its full power in the
description of physical phenomena. Without development of the
geometries of curved spaces, general relativity could not have
emerged. The role of geometry is nowadays being investigated also
within the context of string theory, and especially in the
searches for $M$-theory. The need for suitable generalizations,
such as non commutative geometries is being increasingly
recognized.

It was recognized long time ago [6]  that Clifford algebra
provided a very useful tool for a description of geometry and
physics, containing a lot of room for important generalizations of
the current physical theories. Hence it was suggested in  [3,5]
that every physical quantity is in fact a $ poly-vector$, that
is, a   Clifford number or a  Clifford aggregate. It has turned
out that spinors are the members of left or right minimal ideals
of Clifford algebra, the fact that provided a framework for a
description and a deeper understanding of sypersymmetries, i.e.,
the transformations  relating bosons and fermions.

Moreover, it was shown that the well known Fock-Stueckelberg
theory of relativistic particle [6] can be embedded in the
Clifford algebra of spacetime [3]. Many other fascinating
aspects of Clifford algebra are described in a recent book  [3]
and [5]. A recent overview of Clifford
algebras and their applications is to be found in the book [6].

In addition, the fruitfulness of Clifford algebra showed itself in
the following. A significant body of work was devoted to the
collective dynamics of $p$-branes in terms of area variables [7].
It has been observed [2]  that this has connection to
$C$-space, and also to the branes with variable tension [22] and
wiggly branes [3]. Moreover, using these methods the bosonic
$p$-brane propagator [7] and the logarithmic corrections to the
black hole entropy based on the geometry of Clifford space (in
short $C$-space) was obtained [12].

In previous papers [2,3,11,12] we  proposed a new physical
theory where the arena for physics is no longer the ordinary
spacetime, but a more general manifold of Clifford algebra valued
objects, polyvectors. Such a manifold has been called  a
pan-dimensional continuum  [5]  or $C$-space [2]. The
latter describes on a unified basis the objects of various
dimensionality: not only points, but also closed lines, surfaces,
volumes,.., called 0-loops, 2-loops, 3-loops, etc.. It is a sort
of a $dimension$ category, where the role of functorial maps is
played by C-space transformations. The above geometric objects may
be considered as to corresponding to the well-known physical
objects, namely closed $p$-branes.

The ordinary space-time then is just a subspace of $C$-space. A
``point'' of $C$-space can correspond to any $p$-loop in ordinary
space-time. Rotations in $C$-space transform one point of
$C$-space into another point of $C$-space, and this appears in the
ordinary space as a transformation from a $p$-loop into another
$p'$-loop of different dimensionality $p'$. Technically those
transformations are generalizations of Lorentz transformations to
$C$-space. In that sense, the C-space is roughly speaking a sort
of generalized Penrose-Twistor space.

Furthermore, instead of a flat $C$-space we may consider a curved
$C$-space. Since moving from flat Minkowski space-time to a curved
spacetime had provided us with a very deep insight into the nature
of one of the fundamental interactions, namely gravity, we may
expect that introduction of a curved $C$-space will increase
further our understanding of the other fundamental interactions
and their unification with gravity.

Motivated by these important developments and prospects, we
outline next the programs of the Extended Scale Relativity (ESR)
in C-spaces.

\bigskip

\centerline {\bf Extending Relativity from Minkowski spacetime to
$C$-space}
\bigskip
We embark onto a trip into the extended relativity theory in
C-spaces by a natural  generalization of  the notion of a space-
time interval in Minkwoski space to C-space:

$$ dX^2 = d\Omega^2 + dx_\mu dx^\mu + d x_{\mu\nu} dx^{\mu\nu} + ...
\eqno ( 1 ) $$ The Clifford valued poly-vector:
$$ X = \Omega  I + x^\mu \gamma_\mu +
x^{\mu\nu} \gamma_\mu \wedge \gamma_\nu + ... \eqno ( 2 ) $$
denotes the position in a manifolds, called Clifford space or
$C$-space. If we take differential $ d X$ of $X$ and compute the
scalar product $ d  X *  dX$ we obtain:

$$   d \Sigma^2 =  ( d \Omega)^2 +  \Lambda^{ 2D -– 2 }  dx_\mu dx^\mu +
\Lambda^{ 2D -– 4 }  d x_{\mu\nu}  dx^{\mu\nu } + …..\eqno(3) $$

Here we have introduced  the Planck scale $\Lambda$  since a
length parameter is needed in order to tie objects of different
dimensionality together: 0-loops, 1-loops,..., $p$-loops. Einstein
introduced the speed of light as a universal absolute invariant in
order to ``unite'' space with time (to match units) in the
Minkwoski space interval:
$$ d s^2 =  c^2 d t^2 - d x_i  d x^i . \eqno ( 4 ) $$
A similar unification is needed here to ``unite'' objects of
different dimensions, such as $x^\mu$, $x^{\mu \nu}$, etc...  The
Planck scale then emerges as another universal invariant in
constructing an extended scale relativity theory in C-spaces [2].

To continue along the same path, we consider The analog of Lorentz
transformations in C-spaces transform a poly-vector

$X$ into another poly-vector $X'$ given by $ X' = R X R^{-1} $
with
$$ R = exp~ [ i ( \theta I  + \theta^\mu \gamma_\mu +
\theta^{\mu_1 \mu_2 } \gamma_{\mu_1} \wedge \gamma_{\mu_2 } .....)
] . \eqno ( 5 )    $$ and

$$ R^{-1} = exp~[ - i ( \theta  I + \theta^\nu \gamma_\nu + \theta^{\nu_1
\nu_2 } \gamma_{\nu_1} \wedge \gamma_{\nu_2 } .....) ] .\eqno ( 6
) $$ where the theta parameters:

$$  \theta; \theta^\mu; \theta^{\mu\nu}; ... . \eqno ( 7 ) $$
are the C-space version of the Lorentz rotations/boosts
parameters.

Since a Clifford algebra admits a matrix representation, one can
write the norm of a poly-vectors in terms of the trace operation
as: $ || X ||^2 = {Trace} ~ X^2 $ Hence under C-space Lorentz
transformation the norms of poly-vectors behave like follows:
$$ Trace~  {X'}^2  = Trace ~[ R X^2 R^{-1} ]  =  Trace ~
     [ R R^{-1} X^2 ] = Trace~ X^2 . \eqno ( 8 ) $$
These norms are invariant under C-space Lorentz transformations
due to the cyclic property of the trace operation and $ R R^{-1} =
1 $.

\bigskip

\centerline {\bf Planck scale as the minimum invariant in
Extended Scale Relativity }

\bigskip

Long  time ago L.Nottale   proposed to view the Planck scale as
the absolute minimum invariant (observer independent) scale in
Nature in his formulation of scale relativity [1]. We can apply
this idea to C-spaces by choosing the correct analog of the
Minkowski signature:

$$ || d X || ^2 = d \Sigma^2 = ( d \Omega )^2 [ 1 -  \Lambda^{ 2D - 2 }
{  ( dx_\mu)^2 \over ( d \Omega)^2 }  - \Lambda^{ 2D - 4 } {
(dx_{\mu\nu} )^2 \over ( d \Omega)^2 } - \Lambda^{ 2D - 6 } {
(dx_{\mu\nu\rho} )^2 \over ( d \Omega)^2 }  - ..] $$

$$ || d X || ^2 = d \Sigma^2 = ( d \Omega )^2
[ 1 -  ({ \Lambda \over \lambda_1} )^{ 2D - 2 } - ({ \Lambda \over
\lambda_2} )^{ 2D - 4 }   -    ({ \Lambda \over \lambda_3} )^{ 2D -
6 } - ...…] .  \eqno ( 9 ) $$

where the sequence of $variable$ scales $ \lambda_1, \lambda_2,
\lambda_3, ....$ are  related to the generalized (holographic)
velocities  defined as follows:

$$ {  ( dx_\mu)^2 \over ( d \Omega)^2 }  \equiv  ( V_1)^2 =
( { 1 \over \lambda_1 } )^{ 2D - 2 } . $$

$$ {  ( dx_{\mu\nu} )^2 \over ( d \Omega)^2 }  \equiv  ( V_2)^2 =
( { 1\over \lambda_2 } )^{ 2D - 4 } . $$

$$ {  ( dx_{\mu\nu\rho} )^2 \over ( d \Omega)^2 }  \equiv  ( V_3)^2 =
({ 1 \over \lambda_3 } )^{ 2D - 6 } . . \eqno ( 10 ). $$

. . .

It is clear now that if $ || dX ||^2 \ge 0 $ then the sequence of
$variable$ lengths $ \lambda_n$ cannot be $smaller$ than the
Planck scale $ \Lambda $. This is analogous to a situation with
the Minkoswki interval:

$$ ds^2 = c^2 dt^2 [ 1 - {v^2\over c^2 } ] . \eqno ( 11 ) $$

when it is $ \ge 0 $ if, and only if, the velocity $ v$ does not
exceed the speed of light.  If any of the $\lambda_n$ were smaller
than the Planck scale the C-space interval will become
tachyonic-like $ d \Sigma^2  < 0 $. Photons in C-space are
$tensionless$ branes/loops. Quite analogously one can interpret
the Planck scale  as the postulated minimum universal distance in
nature, not unlike the postulate about the speed of light as the
upper limit on the speed of signal propagation.

What seems remarkable in this scheme of things is the nature of
the signatures and the emergence of two times. One of the latter
is the local mode, a clock, represented by $ t $ and the other
mode is a ``global'' one represented by the volume of the space-time
filling brane $ \Omega $. For more details related to this
Stuckelberg-type parameter and the two modes of time in other
branches of science see [13].

Another immediate application of this       is that one may
consider ``strings'' and ``branes'' in C-spaces as a unifying
description of $ all$ branes of different dimensionality.As we
have already indicated, since spinors are left/right ideals of a
Clifford algebra, a supersymmetry is then naturally incorporated
into this approach as well. In particular, one can have world
volume and target space supersymmetry $simultaneously$ [20].

We hope that the $C$-space ``strings'' and ``branes'' may lead us
towards discovering the physical foundations of string and
M-theory.

In this talk we shall explore several important topics currently
under intensive invesitigation.

\bigskip
\section{\bf Planck-scale Relativity, $\kappa$-deformed Poincare from
C-spaces}
\bigskip
We will briefly summarize the main results of [19] that
allowed  us, among other things,  to derive the Casimirs (masses)
of kappa-deformed Poincare algebras [9,10].  The
relativity in C-spaces is very closely connected to Planck-scale
Relativity [9,10]. Below we will review  how the minimal
length string uncertainty relations can be obtained from  C-spaces
[2].
  The norm of a momentum poly-vector was defined:

$$ P^2 = \pi^2 + p_\mu p^\mu + p_{\mu\nu} p^{\mu\nu} + p_{\mu\nu\rho}
p^{ \mu\nu\rho} + .... =  M^2 \eqno (12 )  $$ Nottale has given
convincing arguments why the notion of $dimension$ is resolution
dependent, and at the Planck scale, the minimum attainable
distance, the dimension becomes singular, that is blows-up. If we
take the dimension at the Planck scale to be infinity, then the
norm $ P^2$ will involve an infinite number of terms since the
degree of a Clifford algebra in $ D$-dim is $ 2^D$.  It is
precisely this infinite series expansion which will reproduce
$all$  the different forms of the Casimir invariant masses
appearing in kappa-deformed Poincare algebras [9,10].

It was discussed recently why there is an infinity of possible
values of the Casimirs invariant $ M^2$ due to an infinite choice
of possible bases. The parameter $ \kappa$ is taken to be equal to
the inverse of the Planck scale. The classical Poincare algebra is
retrieved when $\Lambda = 0$. The kappa-deformed Poincare algebra
does $not$ act in classical Minkwoski spacetime. It acts in a
quantum-deformed spacetime. We conjecture that the natural
deformation of Minkowski spacetime is given by C-space.

The way to generate all the different forms of the Casimirs $ M^2$
is by ``projecting down'' from the $ 2^D$-dim Clifford algebra to
$ D$-dim. One simply  ``slices'' the $ 2^D$-dim mass-shell
hyper-surface in C-space by a $D$-dim one.  This is achieved by
imposing the following constraints on the holographic components
of the polyvector-momentum. In doing so one is explicitly
$breaking$ the poly-dimensional covariance and for this reason one
can obtain an infinity of possible choices for the Casimirs $
M^2$.

To demonstrate this, we impose the following constraints:

$$ p_{\mu\nu} p^{\mu\nu} = a_2 (p_\mu p^\mu)^2 = a_2 p^4. ~~~
p_{\mu\nu\rho} p^{\mu\nu\rho} = a_3 (p_\mu p^\mu)^3 = a_3 p^6 . ~~~
...... \eqno ( 13 )  $$

Upon doing so the norm of the poly-momentum becomes:

$$ P^2 = \sum_n a_n p^{2n}  = M^2 ( 1, a_2, a_3, ..., a_n, ...)  \eqno (14 )
$$
Therefore, by a judicious choice of the coefficients $ a_n $,
and by reinserting the suitable powers of the Planck scale, which
have to be there in order to combine objects of different
dimensions, one can reproduce $all$  the possible Casimirs  in the
form:

$$ M^2 = m^2 [ f ( \Lambda m )] ^2 . ~~~ m^2 \equiv p_\mu p^\mu = p^2.
\eqno ( 15  )  $$ where the functions $ f ( \Lambda m )$ are the
$scaling$ functions with the property that when $ \Lambda  = 0$
then $ f \rightarrow 1 $.
\bigskip

\centerline {\bf The Generalized String Uncertainty Relations }
\bigskip

To illustrate the relevance of poly-vectors, we will summarize our
derivation of the minimal length string uncertainty relations [2].
The canonical quantization in C-space will be given in a future
work. Because  of the holographic variables one cannot naively
impose $ [ x, p] = i\hbar $ due to the effects of the other
components. The units of $[ x_{\mu\nu}, p^{\mu\nu} ] $ are of
$\hbar^2 $ and of higher powers of $\hbar$ for the other
commutators.  To achieve covariance in C-space which reshuffles
objects of different dimensionality, the effective Planck
constant in C-space should be given by a sum of powers of $
\hbar$.

This is not surprising. Classical C-space contains the Planck
scale, which itself depends on $\hbar$. This implies that already
at the classical level, C-space contains the seeds of the quantum
space. At the next level of " quantization, we have an effective
$\hbar$ that comprises all the powers of $\hbar$ induced by the
commutators involving $all$ the holographic variables. In general
one must write down the commutation relations in terms of
polyvector- valued quantities. In particular, the Planck constant
will now be a Clifford number, a polyvector with multiple
components.

The simplest way to infer  the effects of the holographic
coordinates of C-space on the commutation relations is by working
with the effective $ \hbar$  emerging from the ``shadows'' of
C-space. For the relevance of these ``shadows'' of Planck scale
physics to string theory see [18]. The mass-shell condition in
C-space, after imposing the constraints among the holographic
components, yields an effective mass $ M = m f ( \Lambda m ) $.
The generalized De Broglie relations, which are $no$ longer
linear, are [2]:

$$|  P_{effective} |   =  | p | f ( \Lambda m/\hbar ) = \hbar_{effective}
| k| .
~~~ \hbar_{effective} = \hbar f ( \Lambda m/\hbar )  = $$
$$\hbar  \sum  a_n (\Lambda m/\hbar )^{ 2n} . ~~~ m^2 = p^2 = ( \hbar
k)^2. \eqno
( 16 ) $$

Using the effective $ \hbar_{eff}$, the well known relation based
on the Schwartz inequality and the fact that $ | z | \ge | Im z |
$ we obtain:

$$ \delta x^i  \delta p^j  \ge { 1 \over 2 } |  < [ x^i, p^j  ] > |  =
{ \hbar_{effective} \over 2 }  \delta^{ij} . \eqno ( 17 ) $$

Using the relations

$$< p^2 > \ge ( \delta p )^2 .~~~ < p^4 > \ge ( \delta p )^4 . .....
\eqno (18 ) $$

and the series expansion of the effective $ \hbar_{eff}$, we get
for each component (we omit indices for simplicity):

$$ \delta x \delta p \ge { 1\over 2 } \hbar + { a \Lambda^2 \over 2
\hbar } ( \delta p)^2 +............\eqno ( 19 ) $$

This yields the minimal length string uncertainty relations:

$$ \delta x \ge { \hbar \over 2 \delta p } + { a \Lambda^2 \over 2
\hbar } \delta p .....\eqno (20 ) $$ One could include $all$ the
terms in the series expansion and derive a generalized
string/brane uncertainty relation which still retains the minimal
length condition, of the order of the Planck scale [2].

The Physical interpretation  of these uncertainty relations follow
from the extended relativity principle. As we boost the string to
higher energies part of the energy will $always$ be invested into
the string's potential energy, increasing its length in  bits of
Planck scale sizes. This reminds one of  ordinary relativity,
where boosting a massive particle to higher energy increasing its
speed and a part of the energy is also invested into increasing
its mass. In this process the speed of light remains the maximum
attainable speed (it takes an infinite energy  to do so) and in
our scheme the Planck scale is never surpassed. The effects of a
minimal length can be clearly seen in Finsler geometries having
both a maximum four acceleration $ c^2 / \Lambda$ (maximum tidal
forces) and a maximum speed [21]. The Riemannian limit is
reached when the maximum four acceleration goes to infinity; i.e.
the $ \Lambda = 0$ and Finsler geometry " collapses " to
Riemannian one.
\bigskip

\centerline {\bf Effective Lorentz Boosts from C-space Lorentz
Transformations}
\bigskip

We can also show that the effective Lorentz boosts transformations
can be derived from the C-space Lorentz transformations
by a judicious choice of the theta parameters. The effective
boosts  along the $x_1$ direction were obtained in [10] using
the kappa-deformed Poincare algebra:

$$ t' = t~ cosh [ z ( \xi ) ] + x_1 ~ sinh [ z(\xi) ] . ~~~x'_1 = t~
sinh [ z ( \xi ) ] + x_1 ~ cosh [ z ( \xi ) ] . \eqno ( 21) $$
where $ z ( \xi ) $ is the $effective$ boost parameter that
collapses to $ \xi $ when $ \Lambda = 0 $. The effective boost $ z
( \xi )$ ensures that the minimum Planck scale is not surpassed
after the (effective) Lorentz contraction. When one has an
infinite amount of energy, the $ \beta = v/c = 1 $ and the
ordinary boosts are: $ \xi = arctanh ( \beta ) = arctanh ( 1 ) =
\infty$. But the effective boost $ z ( \xi ) = z ( \infty) $ is
$finite$ meaning that boosts $saturate$ at the Planck scale [10]
and the (effective) Lorentz contraction factor  doesn't blow up
which would otherwise have shrunk all lengths to zero.

The C-space Lorentz transformations  of the $ X$ poly-vector can
be written in the most general compact form:

$$ X' = {X'}^N E_N =  e^{ i \theta_A E^A } ( X^M E_M) e^{ - i \theta_B
E^B
} . \eqno (  22 ) $$

where

$$ \theta_A = \{ \theta, \theta_\mu, \theta_{\mu\nu},\theta_{\mu\nu\rho},
.... \} . \eqno ( 23) $$ are the C-space boosts parameters. And
$E^A$ the C-space basis elements.

Performing a Taylor series expansion and taking the scalar product
of both sides of eq. ( 22 ), by $ * E^N$, it can be written in
the form $ X^{'N} \sim {\cal L}^N_M X^M $. It contanins  the
following types of terms:

$$  ( \theta_A \theta^A)^n ( \theta_B \theta^B)^m ~(\theta_C \theta^C) ~
(E_M  * E^N) ~ X^M . $$

$$  ( \theta_A \theta^A)^n ( \theta_B \theta^B)^m ~
\{ \theta_C \theta_D E^C [  E_M, E^D] \}* E^N  ~ X^M . $$

$$  ( \theta_A \theta^A)^n ( \theta_B \theta^B)^m ~ ( E_M * E^N)  X^M .$$

$$  i ( \theta_A \theta^A)^n ( \theta_B \theta^B)^m ~\{ \theta_C E^C E_M \}
*
E^N  ~  X^M . \eqno ( 24 ) $$

Notice that the $odd$ powers of the $\theta$'s are the ones which
contain the imaginary unit.

One can perform a ``dimensional reduction'' from the C-space
Lorentz to an effective Lorentz:

$$ \theta_A E^A \rightarrow \theta^{eff}_{01} \gamma^0 \wedge \gamma^1 \eqno
( 25 ) $$ by imposing the conditions:

$$ \theta_A \theta^A = \sum_k  \alpha_k \xi^k . \eqno ( 26 ) $$
the unknown coefficients $ \alpha_k$ to be be determined later,

   The terms in (24 ) become now:

$$ (  \sum_{m,n }  (\sum_k  \alpha_k \xi^k )^{ m+n  + 1 } ) (   E_M * E^N )
~ X^M   = F_{2} ( \xi )  X^N . \eqno ( 27 ) $$

$$ (  \sum_{m,n }  ( \sum_k \alpha_k \xi^k )^{ m+n  } )
\{   \theta_C \theta_D E^C [  E_M, E^D ]  \} * E^N  ~ X^M   =
F_{1} ( \xi ) {\cal L}^{  ( 2 )N}_M ~ X^M . \eqno ( 28 ) $$

$$ (  \sum_{m,n }  ( \sum_k \alpha_k \xi^k )^{ m+n  } )  (  E_M * E^N )   ~
X^M   = F_{1} ( \xi )  X^N  . \eqno ( 29 ) $$

$$ i  (  \sum_{m,n }  ( \sum_k \alpha_k \xi^k    )^{ m+n  } ) \{ \theta_C
E^C E_M \} * E^N  ~ X^M   = i  F_{1} ( \xi ) {\cal L}^{  (  1 )
N}_M ~ X^M . \eqno ( 30 ) $$

where we have defined:

$$  \sum_{m,n }  ( \sum_k \alpha_k \xi^k )^{ m+n  }  \equiv F_1 ( \xi )
. ~~~ \sum_{m,n }  ( \sum_k \alpha_k \xi^k )^{ m+n + 1  }  \equiv
F_2 ( \xi)
   . \eqno ( 31 ) $$
Notice that both functions of $ \xi $ contain even $and$ odd
powers of $\xi $.

If one sets $all$  the components (except those of the vector $
x^\mu$) of the poly-vector $ X$ to zero, one would arrive at an
$effective$ Lorentz transformation of the form:
$$ x^{'\rho}   = [  e^{ i \theta_A E^A } ( x^\mu \gamma_\mu ) e^{ - i
\theta_B E^B } ] * \gamma^\rho  \equiv L^\rho_\nu  [ \theta,
\theta_\mu, \theta_{\mu\nu}, \theta_{\mu\nu\rho}, .......] ~
x^\nu . \eqno ( 32 )  $$

Since ordinary boosts along the $x_1$ direction (with boost
parameter $\xi$ ) are rotations with an imaginary angle  $ i
\theta_{01} = i \xi $  (  along the $01$ directions) we can see
from the identities:

$$ cos ( i \theta_{01}  ) = cosh ( \theta_{01}  ) = cosh \xi . ~~~
sin ( i \theta_{01}  ) = i sinh ( \theta_{01}  )  = i sinh \xi .
~~~ \theta_{01} \equiv  \xi . \eqno ( 33 ) $$ that one will
retrieve the standard Lorentz Transformations if, and only if, $
\theta_{01}  = \xi $ and if one were to constrain  all of the
other thetas to zero.

In a more general case, this is not so,  and one must include the
contributions of all the other $ \theta_A $. Therefore, when one
uses the constraints imposed on all the C-space theta parameters
given in terms of $ \xi$ by eqs(26), and after defining the
relations for the $two$ functions $ F_1 ( \xi ), F_2 ( \xi )$,
one will have the desired effective Lorentz transformations in
terms of the effective $ z ( \xi ) $ induced from C-space.  The
reason why one has an effective $ z ( \xi )$ is due to the
contributions of all the C-space theta parameters. These
contributions are encoded ( compounded) in the two functions $ F_1
( \xi ), F_2 ( \xi) $, compatible with the fact that we have
$two$ functions $ cosh [ z (\xi ) ]; sinh [ z (\xi ) ] $.

The sought-after equations that define the unknown coefficients $
\alpha_k $   (which determine the constraints among the thetas
and $\xi$) are obtained after using the  double Taylor series
expansion in the following equations:

$$ L^0_0 = L^1_1 = cosh [ z ( \xi ) ] =  \sum_{m, n}   a_n ( b_m \xi^m)^n .
\eqno (34)        $$
$$ L^0_1 = L^1_0 = sinh  [ z ( \xi ) ] =    \sum_{m, n}   c_n ( b_m \xi^m)^n
. \eqno ( 35 )  $$ These are the defining equations that determine
the coefficients $\alpha_k$ in terms of the known coefficients
appearing in the double Taylor series expansion.

In order for the $remaining$ matrix elements $ L^\mu_\nu [ \theta,
\theta_\mu, \theta_{\mu\nu}, ...] $ to be zero one will have to
choose judiciously the thetas to satisfy these vanishing
conditions.; i.e. one will have to choose $most$ of the thetas to
be zero except those whose components contain the $ 01$
directions. In particular,  to ensure that no half-integer powers
of $ \xi $ occur we should choose:

$$ \theta_{ 01 } \sim \xi. ~~~ \theta_{012} = 0. ~~~\theta_{0123} \sim
\xi^2 . ~~~\theta_{01234} = 0. ~~~ \theta_{012345} \sim \xi^3 ..…
\eqno ( 36 ) $$ Otherwise one would have encountered half-integer
valued powers $ \xi^{3/2 },\xi^{5/2},$... $ \xi^{n/2},...$ For
more details see [19].

To conclude: the effective Lorentz boosts are obtained through a
``dimensional reduction'' procedure of the more general C-space
Lorentz transformations
$$ \theta_A E^A \rightarrow \theta^{effective}_{01}  \gamma^0 \wedge
\gamma^1 . \eqno ( 37 ) $$.

For further details about this  derivation and the derivation of
the $nonlinear$  addition law of energy-momenta in particle
collisions in kappa-deformed Minkowski space directly from
C-spaces we refer to [19].  The crux of the
arguments lies in the fact that Planck scale relativity in
C-spaces requires taking the $ D = \infty$ limit.  Because the
conformal group is contained within the Clifford algebra of
space-time [11], the physics of C-space should in principle also
yield the Casimirs for the deformed Weyl conformal algebra of
spacetime [9]. Notice that C-space automatically incorporates
non-commuting objects since poly-vectors, Clifford-valued
matrices, do $not$ commute.

\bigskip

\section{\bf On the geometry of curved $C$-space}

\bigskip

In this section we will summarize the basic results of [11]
that show why the C-space curvature can be written as a sum of
products of the ordinary curvature with torsion. Thus, the analog
of the Einstein-Hilbert action in C-space is given by a higher
derivative gravity with torsion. This result is reminiscent of the
string effective action in curved backgrounds. Since the expansion
is given in powers of the Planck scale .

Let us now consider a curved $C$-space. A basis in $C$-space is
given by

$$  E_A = \{  \gamma_\mu, \gamma_\mu \wedge \gamma_\nu, \gamma_\mu
\wedge \gamma_\nu \wedge \gamma_\rho,... \}  .$$ where in an
$r$-vector $\gamma_{\mu_1} \wedge \gamma_{\mu_2} \wedge ... \wedge
\gamma_{\mu_r}$ we take the indices so that $\mu_1 < \mu_2 <
...<\mu_r$. An element of $C$-space is a Clifford number,  or
$Polyvector$  written in the form

   $$  X = X^A E_A = s I  + x^\mu \gamma_\mu +
        x^{\mu \nu} \gamma_\mu \wedge \gamma_\nu + ... \eqno ( 38 ) $$
A $C$-space is parameterized not only by 1-vector coordinates
$x^\mu$ but also by the 2-vector coordinates $x^{\mu \nu}$,
3-vector coordinates $x^{\mu \nu \rho }$, etc., called also
holographic coordinates, since they describe the holographic
projections of 1-lines, 2-loops, 3-loops, etc., onto the
coordinate planes. By $p$-loop we mean a closed $p$-brane; in
particular, a 1-loop is closed string, a $ 2$-loop is a closed
membrane, etc....

In order to avoid using the powers of the Planck scale length
parameter $\Lambda$ in the expansion of the poly-vector $X$ we use
the dilatationally invariant units in which $\Lambda$ is set to 1
[3]. The Planck scale in general is given in terms of the
Newton constant: $ \Lambda = (G_D)^{ 1/ D -2} $, in units of $
\hbar = c = 1 $. If we imagine performing an analytical
continuation of the dimension from $ D = 2 $ all the way to $ D =
\infty$, it is clear that that in order to have the Planck scale
as a universal invariant,  with the same value in all dimensions,
the value $ \Lambda = 1 $ is compatible with the choice of $ G_D =
1 $ (for all dimensions).  Hence the extended relativity
principle in C-space admits the natural system of units $ \hbar =
c = G = \Lambda = 1 $ which will set the Planck temperature and
Boltzmann constant to be also $ T_P = K_B = 1 $. This $unifying$
picture of setting all the fundamental constants of Nature to  $1$
was advocated by Wheeler long time ago by suggesting that
information theory may lie at the core of things.

In a flat $C$-space the basis vectors $E^A$ are constants. In a
$curved$ $C$-space this is no longer true. Each $E_A$ is a
function of the $C$-space coordinates $ X^A = \{  s, x^\mu,
\sigma^{\mu \nu}, ... \} $  which include scalar, vector,
bivector,..., $r$-vector,..., coordinates.

Now we define the connection ${\tilde \Gamma}_{AB}^C$ in $C$-space
according to $ \partial_A  E_B = {\Gamma}_{AB}^C E_C$. The
$C$-space curvature is thus defined in the geometric calculus
notation:

   $$  {{\cal R}_{ABC}}^D =  ([\partial_ A, \partial _B] E_C) * E^D .
\eqno ( 39 ) $$

The `star' means the $ scalar$  product  between two poly-vectors
$A$ and $B$, defined as $ A*B = < A  | B >_s  $  'where $s$ means
'the scalar part' of the geometric Clifford product $AB$.

Next we shall provide the relation for curvature and see how it is
related to the curvature of the ordinary space.  We will present
the basic formulae and refer  all the details of the derivation to
the references  [11].  We found, in particular that:

$$ { \partial \gamma_\mu \over \partial x^\nu } =
\Gamma^{\alpha}_{\nu\mu } \gamma_\alpha. ~~~ { \partial \gamma_\mu
\over \partial x^{\alpha \beta} } = R_{\alpha \beta \mu}^{ \rho}
\gamma_\rho . \eqno ( 40 ) $$

For an arbitrary poly-vector
$$ V = V^A E_A = v + v^\mu \gamma_\mu + v^{\mu\nu} \gamma_\mu \wedge
\gamma_\nu + ...\eqno ( 41 ) $$

we have the covariant derivatives:

$$ { D V ^A \over D X^B } = {\partial V^A \over \partial X^B } +
\Gamma^A_{BC} V^C. ~~~ { D E^A \over D X^B} = 0 .  . \eqno ( 42 )
$$

$$ { D v \over D x ^{\mu \nu} } = [ D_\mu, D_\nu ] v = { \partial v \over
\partial x^{\mu\nu} } = K^\rho_{\mu\nu} \partial_\rho v.  \eqno ( 43 ) $$

$$ { D v^\alpha  \over D x^{\mu\nu} } = [ D_\mu, D_\nu ] v^\alpha =
R_{\mu \nu\rho }^\alpha  v^\rho  + K_{\mu\nu}^{\rho} D_\rho
v^\alpha . \eqno ( 44 )   $$

using these relations in the basic formula for the curvature:

$$ {{\cal R}_{ABC}}^D =  ([\partial_ A, \partial _B] E_C) * E^D . \eqno
(45  ) $$

we have for a particular example the poly-vector valued
multi-indices:

$$ A = [ \mu\nu ] . ~~~ B = [ \alpha \beta ] . ~~~ C = \tau . ~~~ D =
\delta \eqno ( 46 ) $$

$$  ( [  { \partial \over \partial \sigma^{\mu\nu} }, { \partial \over
\partial \sigma^{ \alpha \beta } } ] \gamma_\tau  ) . \gamma^\delta =
{ \cal R }_{ [ \mu\nu ] [ \alpha \beta ] \tau }^\delta  = $$

$$  R_{\mu\nu \tau }^\rho R_{\alpha \beta \rho}^\delta
- R_{\alpha \beta \tau }^\rho R_{\mu \nu  \rho}^\delta . \eqno (
47 ) $$

We can see now how one gets the product of two usual
curvature tensors. We can proceed in analogous way to calculate
the other components of the C-space curvature $ {\cal R}_{ABC}^D $
and find that these contain $higher$ powers of the curvature in an
ordinary space-time.  After performing the appropriate
contractions, we can see that the scalar curvature in C-space will
contain the sums of products of the ordinary curvature tensors. In
general one has also contributions from a non-vanishing torsion.
This resembles the results based on non-linear quantum sigma
models and used to evaluate the string effective action in curved
backgrounds as an expansion in higher derivative terms (higher
power of the curvature) [11].

One of the most important consequences of this result is that a
$flat$ C-space can be curved from  the ordinary space-time point
of view.  If one takes the symmetric spaces, like de Sittter and
Anti de Sitter, where the curvature tensors are suitable
multiples of the scalar  curvature, we see that the C-space scalar
curvature is given as a sum of powers of the space-time scalar: $
{ \cal R } = \sum a_n R^n $. Clearly, if $ R = 0 $ then $ { \cal R
} = 0.$ Inversely, $ { \cal R } = 0$ may yield besides the trivial
solution $ R = 0 $ other $nonvanishing$ values for $ R $, such as
the solutions of a polynomial equation. Tele-paralellism theories
of gravity have vanishing curvature but non-vanishing torsion.

Based on this fact, the logarithmic corrections in any dimension
to black hole entropy  were found with the help of a quantum
p-loop harmonic oscillator       in flat C-space [12]. The
logarithm of the degeneracy of quantum states yielded the standard
Bekenstein-Hawking area-entropy relation plus the logarithmic
corrections expressed in terms of Clifford-bits or true quanta
$of$ spacetime: the holographic areas, holographic volumes,
...were all quantized in Planck scale units.

Furthermore, the expression for the Schwrazschild radius and
Hawking temperature, in any dimension, were also found using
these methods [12]. The open problem was to derive the
Schwarzschild metric from the long distance limit of the
$condensation $ of the holographic quanta of area bits, volume
bits, ... The most important result in [12] was that C-space
methods  appear to indicate naturally the origins of the
thermodynamic properties of black holes and why string theory
contains gravity.

\bigskip

\section {\bf Cosmological Implications and Further topics }

\bigskip

To end this talk, I sketch the cosmological and other physical
implications of the extended scale relativity in C-spaces.

$\bullet $  Cosmological constant. Flat C-space versus non-flat
Riemann.

The cosmological " constant " is just one component of the
momentum polyvector, the Fourier conjugate to the position
polyvector $ X$. The first component was the volume $ \Omega$  of
the spacetime filling p-loop/brane, $ p+ 1 = D $. Its Fourier
dual is the cosmological " constant, the
vacuum energy density. Clearly, since it is just $a$ component of
a polyvector, the cosmological " constant " is $not$ a constant in
C-spaces.  Secondly, we have discussed above why a $flat$ C-space
may not imply a flat ordinary spacetime. Hence, we hope that
C-space relativity may bring about the key to solve this problem.
Perhaps, the problem was ill-posed due to the fact that the
cosmological ``constant'' may not be a true constant. It is clearly
not a constant in C-space !

$\bullet$ Two times. Universal time arrow. Why universe may
expand forever.

We have also discussed why the volume $ \Omega$  of the spacetime
filling brane plays the role of the " global " time mode, the
Stuckelberg parameter [3] and its relation to the two modes of
time given in  [13]. It is possible that the fundamental
constants, like the fine structure constant,themselves depend  on
this global mode time parameter $ \Omega $.  As the universe
expands, it does according to this universal time arrow. Since
this arrow points " forward" then one should expect that $ \Omega$
will increase and not surprisingly the universe will expand
forever.

On the variable fine structure, we know from the above results
that C-space demands an effective $ \hbar$ given by a sum of
powers of $ \Lambda $. If one uses $ \alpha = e^2 / \hbar_{eff}  c
$ then one should expect corrections to the ordinary value.
However, this not new since  results of the renormalization group
program require running coupling constants. What is new is the
dependence on all " constants " on the global mode of time
$\Omega$. This idea is more compatible with Dirac's picture of the
changing values with time of the fundamental constants in Nature.

$ \bullet$  Yang-Mills interactions. Some preliminary results on
how to incorporate Wilson loops in C-spaces appeared in [15].
The construction of fiber bundles over p-loop spaces seems to be a
challenging problem.

$\bullet$ Duality of small/large scales. Nottale had envisioned
this duality long ago and wrote down the analog of scale
relativistic transformations involving an upper impassible scale,
another universal invariant, the dual picture of the minimum
Planck scale.  This was his proposal for the resolution of the
cosmological constant problem: it is meaningless to compare the
vacuum energy at two such separate scales, the Planck versus the
Hubble regime, without including the scale relativistic
corrections. This accounted perfectly for the $ 10^{ 60}$ factors.

$\bullet $ Quantum Clifford algebras. One can " deform" C-spaces
by using q-Clifford algebras,  like braided Hopf quantum Clifford
algebras [16]. For an extensive report on q-spins see [17].
For other relevant work on deformed Poincare algebras, quantum
groups, in the construction of q-deformed Lagrangians for gravity
see [14].

$\bullet $ On four as the average dimension of the world.

Since Planck scale relativiy involves an infinite number of
dimensions one will immediately wonder why we perceive four
dimensions. In [23] it was shown that the average dimension of
a family  of spheres of arbitrary dimensions was close to $ 4 +
\phi^3 = 4.236 $, where $ \phi$ is the Golden Mean $ 0.618 ..$.

\bigskip

\centerline {Acknowldegement}

We are indebted to Alex Granik,  Matej Pavsic and Laurent Nottale
for many discussions and critical comments. To Jorge Mahecha for
his valuable help, to Meredith  Bowers for her kind hospitality
in Santa Barbara where this work was completed and to Metod Saniga
for his kind invitation to the NATO Advanced Workshop on the
Nature of Time: Geometry, Physics and Perception. Slovakia, May 2002.
\bigskip

\centerline {References}
\bigskip

1- L. Nottale: La Relativite dans tous ses Etats " Hachette
Lit. Paris 1999.

" Fractal Spacetime and Microphysics, Towards Scale Relativity. World
Scientific. Singapore, 1992.

2-C. Castro: Chaos, Solitons and Fractals {\bf 11} (2000) 1663.

hep-th/0001134

Foundations of Physics {\bf 30} (2000) 1301.  hep-th/0001023

Chaos, Solitons and Fractals {\bf 11} (2000) 1721.

hep-th/9912113

Chaos, Solitons and Fractals {\bf 12} (2001) 1585.

physics/0011040

``The search for the origins of M theory ....'' hep-th/9809102

3-M.Pavsic: The Landscape of Theoretical  Physics:

A Global View " Kluwer, Dordrecht 1993.

" Clifford algebra based polydimensional Relativity and

Relativistic Dynamics "

Talk presented at the IARD Conference in Tel Aviv, June 2000.

Foundations of Physics {\bf 31} (2001) 1185.  hep-th/0011216.

Phys. Let {\bf A 242} (1998) 187.

Nuovo Cimento {\bf A 110} (1997) 369.

4-J. Fanchi " Parametrized Relativistic Quantum Theory.

Kluwer, Dordrecht 1993.

5-W. Pezzaglia: ``Physical Applications of a Generalized
Geometric Calculus'' gr-qc/9710027.

6-D. Hestenes: Spacetime Algebra "  Gordon and Breach, New
York, 1996.

D. Hestens and G. Sobcyk: Clifford Algebra to Geometric
Calculus " D. Reidel Publishing Company, Dordrecht, 1984.

" Clifford Algebras and their applications in Mathematical Physics
" Vol 1: Algebras and Physics. eds by R. Ablamowicz, B. Fauser.

Vol 2: Clifford analysis. eds by J. Ryan, W. Sprosig.
Birkhauser, Boston 2000.

P. Lounesto: Clifford Algebras and Spinors. Cambridge
University Press. 1997.

7- S. Ansoldi, A. Aurilia, E. Spallucci: Chaos, Solitons and
Fractals {\bf 10} (2-3) (1999).

8- S. Ansoldi, A. Aurilia, C. Castro, E. Spallucci: Phys. Rev.
{\bf D 64 } 026003 (2001) hep-th/0105027.

9-J. Lukierski, A. Nowicki: Double Special Relativity versus
kappa-deformed Relativistic dynamics " hep-th/0203065.

J. Lukierski, V. Lyakhovsky, M. Mozrzymas: kappa-deformations
of D = 4 Weyl and conformal symmetries " hep-th/0203182.

S.Majid, H. Ruegg: Phys. Let {\bf B 334} (1994) 348.

J. Lukierski, H. Ruegg, W. Zakrzewski: Ann. Phys. {\bf 243 }
(1995) 90.

J. Lukierski, A. Nowicki, H, Ruegg, V. Tolstoy: Phys. Lett {\bf B
264} (1991) 331.

10-J. Kowalski-Glikman, S. Nowak: Doubly special Relativity
theories as different bases of  kappa-Poincare algebras. hep-th/0203040.

N. Bruno, G. Amelino-Camelia, J. Kowalski: Deformed boosts
transformations that saturate at the Planck scale " hep-th/0107039

G. Amelino-Camelia: Int. J. Mod. Phys {\bf D 11} (2002) 35.
gr-qc/0012051

G. Amelino-Camelia: Phys. Lett {\bf B 510} (2001) 255.

A. Granik: A comment on the work of Bruno-Amelino-Camelia and
Kowalski " physics/0108050.

11-C. Castro, M. Pavsic: Higher Derivative Gravity and Torsion
from the Geometry of C-spaces " hep-th/0110079

C. Castro, M. Pavsic: The Clifford algebra of Spacetime and the Conformal
Group "

12-C. Castro, A. Granik: Extended Scale Relativity, p-loop
harmonic oscillator and logarithmic corrections to the black hole entropy "
physics/0009088.

C. Castro: Jour. of Entropy {\bf 3} (2001) 12-26.

13-D. Chakalov: " Two modes of Time: Bicausality " Talk to be
presented at the NATO advance workshop on the Nature of Time: Geometry,
Physics and Perception. Slovakia, May 2002.

14-L. Castellani: The Lagrangian of q-Poincare Gravity "
hep-th/940233

" Differential Caluculus on $ISO_q (N) $, Quantum Poincare
Algebra and q-Gravity " hep-th/9312179.

15-C. Castro: On Wilson Loops and Confinement without
Supersymmetry from Composite Antisymmetric Tensor Field Theories:

hep-th/0204182.

16-Z. Osiewicz: Clifford Hopf Algebra and bi-universal Hopf
algebra " q-qlg/9709016

17-C. Blochmann " Spin representations of the Poincare Algebra "
Ph. D Thesis math.QA/0110029.

18- K. Dienes, A. Mafi: Phys. Rev. Let {\bf 88} (11) (2002)
111602

19- C. Castro, A. Granik: " Planck scale Relativity and variable
fine structure from C-space "To appear.

20- A. Aurilia, C. Castro, M. Pavsic, E. Spallucci: To appear.

21-H. Brandt: Chaos, Solitons and Fractals {\bf 10} (2-3)
(1999) 267.

22-E. Guendelman: Class. Quant. Grav {\bf 17} (2000) 3673.
hep-th/0005041.

23-C. Castro, A. Granik: Chaos, Solitons and Fractals {\bf 12}
(10) (2001) 1793.

\end{document}